\documentstyle[12pt,epsf]{article}

\def\starup#1{\mbox{$\raise1.8ex\hbox{$*$} \kern-.7em#1$}}

\title{Bounds on scalar leptoquark masses \\
       from S, T, U parameters in the minimal  \\ 
       four-color quark-lepton symmetry model }
 
\author{A.D.~Smirnov\thanks{E-mail: asmirnov@univ.uniyar.ac.ru}\\
{\small\it Division of Theoretical Physics, Department of Physics,}\\
{\small\it Yaroslavl State University, Sovietskaya 14,}\\
{\small\it 150000 Yaroslavl, Russia.}}

\date{}

\begin{document}

\maketitle

\begin{abstract}

The contributions into radiative correction parameters $ S, T, U $ 
from the scalar leptoquarks are calculated in the minimal 
gauge model with the four-color quark-lepton symmetry.
It is shown that the contributions into $T$ and $U$ from the 
scalar leptoquark doublets are not positive definite.
Using the current experimental data on $S, T, U $ 
the bounds on the scalar leptoquark masses are obtained.  
In particular, the existence of the relatively light 
scalar leptoquark with electric charge~2/3  is shown 
to be compatible with the current experimental 
data on $S, T, U$. 

\end{abstract}

\newpage

The search for a new physics beyond the Standard Model ( SM ) is now 
one of the aims of the high energy physics.One of the possible variants
of the new physics beyond the SM can turn out to be the variant induced 
by the possible four-color symmetry \cite{PS} between quarks and leptons.
The investigation of the possible manifestations of this symmetry at 
attainable energies is of a certain interest now.

The main feature of the four-color  gauge symmetry is the prediction 
of the vector leptoquarks the masses of which are expected to be of 
about the mass scale $M_c$ of the four-color symmetry breaking. 
There are many models dealing with the four-color gauge symmetry now. 
Some of them are induced by the grand unification ideas and regard 
the four-color symmetry  as an intermediate stage of the symmetry 
breaking. The mass scale $M_c$ in these models is usually very high. 
For instance, in $SO(10)$ model it is of about $M_c\sim 10^{12}\,\,GeV$ 
\cite{EF} but it can be essentially lowered up to $M_c\sim 10^5 - 10^6 
\,\,GeV$ by appropriate scheme of the symmetry breaking \cite{SS}. 
The other models \cite{AD1,AD2,V,RF,BL} regard the four-color symmetry 
as a primary symmetry with the mass scale $M_c$ determined mainly 
by the low energy experimental data. At such approach  
the lower limit on $M_c$ can be about $1000 \, TeV$ 
\cite{V} or about 100 $TeV$ or slightly less \cite{AD1,AD2} 
or even can be lowered up to $1 \, TeV$ by the appropriate 
arrangement of the fermions in the $SU(4)$- multiplets \cite{RF,BL}.
     
It should be noted that 
the four-color symmetry can manifest itself not only by the vector 
leptoquark physics but also due to the scalar leptoquarks. Because of their 
$SU(2)-$ doublet structure the scalar leptoquarks can manifest themselves 
in the oblique radiative corrections, in particular, affecting the 
$S, T, U$ parameters of Peskin and Takeuchi \cite{PT}. It is interesting 
to know the effect of the scalar leptoquarks on $S, T, U$ in some 
simple gauge model predicting these particles. 

In this work the contributions into $S, T, U$ parameters from the scalar 
leptoquarks are calculated using the minimal quark-lepton symmetry model 
( MQLS-model ) \cite{AD1} as a minimal extension of the SM containing the 
four-color quark-lepton symmetry. Taking the current experimental values 
of $S, T, U$ into account the bounds on scalar leptoquark masses are obtained 
and discussed. 

The MQLS-model to be used here is based on the
$SU_V(4) \times SU_L(2)\times U_R(1)$-group 
and predicts the new 
gauge particles ( vector leptoquarks 
$V{\alpha \mu}^\pm$ with electric charge ${\pm}$2/3 and an extra neutral 
$Z'-$boson ) as well as the new scalar ones \cite{AD1,AD2}.

The scalar sector of the model contains in general the four multiplets 
\begin{eqnarray}
(4,1,1):   \Phi^{(1)} & =  &
\left ( \begin{array}{c} S_{\alpha}^{(1)}    \\
  \frac {\eta_1 + \chi^{(1)}+i\omega^{(1)}}{\sqrt 2} \end{array} \right ) , 
\nonumber \\
(1,2,1):   \Phi_a^{(2)} & = & \delta_{a2}\frac{\eta_2}{\sqrt 2} + \phi_a^{(2)}, 
\nonumber \\
(15,2,1):  \Phi_a^{(3)} & = & \left( 
\begin{array}{cc}
 (F_a)_{\alpha\beta} & S_{a\alpha}^{(+)} \\
 S_{a\alpha}^{(-)}   &  0  
\end{array} \right) +  \Phi_{15,a}^{(3)}t_{15},
\nonumber \\
(15,1,0):  \Phi^{(4)} & = & \left( 
\begin{array}{cc}
 F_{\alpha\beta}^{(4)} & \frac 1{\sqrt 2}S_{\alpha}^{(4)} \\
\starup{S_\alpha^{(4)}}   &  0  
\end{array} \right) + (\eta_4 + \chi^{(4)})t_{15} ,
\nonumber
\end{eqnarray}
transforming according to the  (4,1,1)-,(1,2,1)-,(15,2,1)-,(15,1,0)- 
representations of the
$SU_V(4) \times SU_L(2)\times U_R(1)$-group 
respectively. Here 
$
 \Phi_{15,a}^{(3)}= \delta_{a2}\eta_3 + \phi_{15,a}^{(3)}, 
$ 
$\eta_1$, $\eta_2$, $\eta_3$, $\eta_4$ are the vacuum expectation
values, $t_{15}$ is the 15-th generator of $SU_V(4)$-group,$ a=1,2
$ is the $SU_L(2)$ index and $ 
\alpha,\beta=1,2,3$ are the $SU_c(3)$ color indexes.

As regards the $S, T, U$ parameters the most intersting of these 
multiplets is the muliplet $\Phi^{(3)}$ which has been introduced 
to split the masses of quarks and leptons.This multiplet contains 
fifteen doublets which can be arranged into two scalar leptoquark 
doublets $ S_{a\alpha}^{(+)}$ and $S_{a\alpha}^{(-)}$ with the SM hypercharge 
$Y^{(SM)}$=7/3 and -1/3 respectively,eight scalar gluon doublets 
$F_{ja}$, j=1,2,...8 with $Y^{(SM)}$=1 and the doublet 
$\Phi_{15,a}^{(3)}$ which in admixture with $\Phi_a^{(2)}$
gives the SM doublet
\begin{eqnarray}
 \Phi^{(SM)} = 
\left ( \begin{array}{c} \Phi_1^{(SM)}  \\
  \frac {\eta + \chi_1 + i\omega_1}{\sqrt 2} \end{array} \right ) 
\nonumber
\end{eqnarray}
with SM VEV $\eta=\sqrt{{\eta_2}^2+{\eta_3}^2}=(\sqrt 2 G_F)^{-\frac 12}
\approx 250\ GeV$ and an additional scalar doublet $\Phi'$.

All these scalar doublets contribute into $S, T, U$ pararmeters.
We have calculated these contributions neglecting the effect 
of the small $Z-Z'$-mixing.

The contributions into $S, T, U$ from the doublets $\Phi_a'$ and 
$F_{ja}$ have the usual form of those from a single scalar doublet 
${\Phi}$ and can be written as
 
\begin{eqnarray}
S^{(\Phi)} = -k_{\Phi} \frac { Y^{SM}}{12 \pi} 
\ln{ \frac {m_1^2}{m_2^2}},   \label{eq:sp} \\
T^{(\Phi)} = k_{\Phi} \frac {1}{16\pi{c_W}^2{s_W}^2{m_Z}^2} f_1(m_1,m_2),
\label{eq:tp}  \\ 
U^{(\Phi)} = k_{\Phi} \frac {1}{12 \pi} f_2(m_1,m_2),
\label{eq:up} 
 \end{eqnarray}
where
\begin{eqnarray}
f_1(m_1,m_2) & = & {m_1}^2 + {m_2}^2 - \frac {2m_1^2m_2^2}{{m_1}^2 - {m_2}^2}
\ln{ \frac {m_1^2}{m_2^2}}, 
\label{eq:f1} \\
f_2(m_1,m_2) & = & - \frac {5{m_1}^4 + 5{m_2}^4 - 22{m_1}^2{m_2}^2 }
{3({m_1}^2 - {m_2}^2)^2}
\nonumber \\
& + & \frac {{m_1}^6 - 3{m_1}^4{m_2}^2 - 3{m_1}^2{m_2}^4 + {m_2}^6}
{({m_1}^2 - {m_2}^2)^3}
\ln{ \frac {m_1^2}{m_2^2}}
\label{eq:f2}
\end{eqnarray}
and $m_1,m_2$ are the masses of the up and down components of the doublet
$\Phi$. Here 
$k_{\Phi'}$=1, $m_a=m_{\Phi_a'}$ and 
$k_{F}$=8, $m_a=m_{F_{ja}}$ for multiplets ${\Phi}'$ and $F$ respectively.

The contributions into $S, T, U$ from the scalar leptoquarks are more 
complicated mainly owing to the existence of the Goldstone mode 
$S_0$ among the scalar leptoquarks fields.Indeed, in addition to the 
SM Goldstone modes $\Phi_1^{(SM)}$ and $\omega_1$ the model has the 
Goldstone modes $\omega^{(1)}$ and
\begin{eqnarray}
S_0 = \left \lbrack \frac{\eta_1}2S^{(1)} + \sqrt{ \frac 23}(\eta_3 
\frac{S_2^{(+)}+\starup {S_2^{(-)}}}{\sqrt 2} + \eta_4S^{(4)}) \right \rbrack
/ \sqrt{ \frac {{\eta_1}^2}4 + \frac 23({\eta_3}^2+{\eta_4}^2)}  \,\,\,. 
\nonumber
\end{eqnarray} 

Because the Goldstone field $S_0$ is constructed from the fields
 $S^{(1)}$, $S_2^{(+)}$, $\starup {S_2^{(-)}}$, $S^{(4)}$ with 
electric charge 2/3 the mixing between
these fields is required and they must be expressed as the superpositions 
of $S_0$ and of the orthogonal to it the mass eigenstate leptoqurk fields 
$S_1$, $S_2$, $S_3$. In general these superpositions can be written as
\begin{eqnarray}
S_2^{(+)} & = & \sum_k C_k^{(+)}S_k, \  \ 
\starup {S_2^{(-)}}= \sum_k C_k^{(-)}S_k, \label{eq:sup} \\
S^{(1)}  & =   &\sum_k C_k^{(1)}S_k, \  \
S^{(4)}= \sum_k C_k^{(4)}S_k, 
\nonumber
\end{eqnarray}
where $C_k^{(\pm)}$, $C_k^{(1)}$, $C_k^{(4)}$
are the elements of a unitary mixing matrix, $k=0,1,2,3$.

In the unitary gauge the Goldstone fields are eliminated
$$
\Phi_1^{(S)}=0, \omega_1=0, \omega^{(1)}=0, S_0=0
$$
and the physical scalar leptoquarks conrtibuting into $S, T, U$ 
are two colored thriplets of up scalar leptoquarks 
$S_{1\alpha}^{(+)}, S_{1\alpha}^{(-)}$ with electric charge 5/3 and 1/3 
respectively and three scalar leptoquarks $S_{k\alpha},k=1,2,3$  
with electric charge~2/3. The contribution into $S, T, U$ from 
the SM Higgs particle $\chi_1$ has the usual form and the other scalar 
fields   
$\chi^{(1)}$,$\chi^{(4)}$ and $F_j^{(4)}$, $j=1,2,...8$ as well as 
the new gauge fields $Z'$ and $V_{\alpha}^{(\pm)}$ do not contribute 
into $S, T, U$.

The self energy diagramms contributing into $S, T, U$ from the leptoquarks 
are shown in Fig.1.
The coupling constants 
$g_{XS_kS_l}$, 
$g_{XS_1^{(\pm)}S_1^{(\pm)}}$, $g_{WS_1^{(\pm)}S_k}$, 
$g_{XVS_k}$, $g_{WVS_1^{(\pm)}}$ 
($X$ is the photon $A$ and $Z$-boson) are defined by the model and
depend on the gauge coupling constants and on the matrix elements
of the mixing matrix with   
$g_{ZVS_k}$,  $g_{WVS_1^{(\pm)}}$ depending also on the VEV
$\eta_3$ ( $g_{AVS_k}=0$ in the unitary gauge).

We have calculated the contributions $S^{(LQ)}$, $T^{(LQ)}$,
$U^{(LQ)}$ into $S$-, $T$-, $U$- parameters from the leptoquarks. 
The result can be presented in the form  

\begin{eqnarray}
&& S^{(LQ)} = \frac {n_c}{12 \pi} \Bigg \{ -\sum_{+,-} \sum_k 
\vert C_k^{(\pm)}\vert^2
Y_\pm^{SM} \ln\frac {m_\pm^2}{m_k^2} + \frac 12 \sum_k \sum_l
B_{kl}f_2(m_k,m_l) 
\label{eq:slq}\\
&& +  \xi^2 \bigg \lbrack \sum_k \vert C_{1k} \vert^2
\left ( -12 \frac {m_V^2f_1(m_k,m_V)}{(m_k^2-m_V^2)^2} 
+ f_2(m_V,m_k)\right )  
 + \frac 12 \sum_{+,-}
 \ln\frac {m_V^2}{m_\pm^2}-
\frac 23 \ln\frac {m_+^2}{m_-^2}
\bigg \rbrack \Bigg \},
 \nonumber \\
&& T^{(LQ)} = \frac {n_c}{16 \pi s_W^2c_W^2m_Z^2} \Bigg \{ \sum_{+,-} \sum_k 
\vert C_k^{(\pm)}\vert^2
f_1(m_\pm,m_k) - \frac 12 \sum_k \sum_l
B_{kl}f_1(m_k,m_l) 
\label{eq:tlq}\\
&& +  \xi^2 \sum_{+,-} \sum_k \vert C_{2k} \vert^2 \left \lbrack
\frac 12 \left (  f_1(m_\pm,m_V) - f_1(m_k,m_V)  \right ) + 
 4m_V^2f_3(m_\pm,m_k;m_V)
\right \rbrack \Bigg \},
 \nonumber \\
&& U^{(LQ)} = \frac {n_c}{12 \pi} \Bigg \{ \sum_{+,-} \sum_k 
\vert C_k^{(\pm)}\vert^2
f_2(m_\pm,m_k) - \frac 12 \sum_k \sum_l
B_{kl}f_2(m_k,m_l) 
\label{eq:ulq}\\
&& +  \xi^2 \sum_{+,-} \sum_k \vert C_{2k} \vert^2 \bigg \lbrack 
\frac12\left(f_2(m_\pm,m_V)-f_2(m_k,m_V)\right) 
\nonumber \\
&& - 6m_V^2\left(\frac { f_1(m_\pm,m_V)}{(m_\pm^2-m_V^2)^2}-
\frac{f_1(m_k,m_v)}{(m_k^2-m_V^2)^2}\right)\bigg \rbrack\Bigg\}, 
 \nonumber
\end{eqnarray}
where $f_1(m_1,m_2)$, $f_2(m_1,m_2)$ are defined by eqs.(\ref{eq:f1},\ref{eq:f2})
and 
$$
f_3(m_1,m_2;m_V)= \frac {m_1^2m_2^2 \ln (m_1^2/m_2^2) + 
m_V^2(-m_1^2 \ln (m_1^2/m_V^2)+ m_2^2 \ln (m_2^2/m_V^2))}
{(m_1^2-m_V^2)(m_2^2-m_V^2)}.
$$
Here
$$ 
B_{kl}= \vert \ \starup {C_k^{(+)}}C_l^{(+)}-\starup {C_k^{(-)}}C_l^{(-)}
\vert^2, \ C_k^{(\pm)}= \frac {1}{\sqrt 2}(\sqrt{1-\xi^2}C_{1k} \pm C_{2k}),
\\ k,l=1,2,3
$$ 
$$
\xi^2=\frac23 \eta_3^2/(\eta_1^2/4 + \frac23(\eta_3^2 + \eta_4^2))=
 \frac23g_4^2\eta_3^2/m_V^2,
$$ 
$Y_\pm^{SM}=1\pm 4/3,  n_c=3,  m_k=m_{S_k},  m_{\pm}=m_{S_1^{(\pm)}}$,  
 $g_4$ is the $SU_V(4)$ - coupling constant. 
$C_{1k}, C_{2k}$ are two unit mutually orthogonal complex vectors.
In general, the vectors $C_{1k}, C_{2k}$ can be parametrized by means of three 
mixing angles and three phases, for example, as
\begin{eqnarray}
C_{1k}= \{c_{13}c_{12}, \,\,\,\, c_{13}s_{12}e^{i\delta_{12}}, 
\,\,\,\, s_{13}e^{i\delta_{13}} \}, \nonumber \\
C_{2k}= \{-c_{23}s_{12}e^{-i\delta{12}}- s_{23}s_{13}c_{12}e^{i(-\delta_{13}+
\delta_{23})}, \nonumber \\
c_{23}c_{12}- s_{23}s_{13}s_{12}e^{i(-\delta_{13}+\delta_{12}+\delta_{23})},
\,\,\,\, s_{23}c_{13}e^{i\delta_{23}} \}, 
\nonumber 
\end{eqnarray}
where $c_{ij}=\cos \theta_{ij}, s_{ij}=\sin \theta_{ij},
\theta_{ij}, \delta_{ij} $ are the mixing angles and phases. 
     
The contributions (\ref{eq:slq})-(\ref{eq:ulq}) into $S, T, U$ from 
the scalar leptoquarks 
differ essentially from those arising from ordinary scalar doublets 
and having the form (\ref{eq:sp})-(\ref{eq:up}). 
Firstly, the contributions (\ref{eq:tlq}), (\ref{eq:ulq}) from the scalar 
leptoquarks  into $T$ 
and $U$  are not positive definite due to the mixing between the scalar 
leptoquarks with electric charge 2/3. The mixing of the components of 
two scalar doublets as a possible mechanism for obtaining the negative 
$T$ and $U$ was pointed out in Ref.\cite{LL}. But in our case the 
$S_1-S_2-S_3$ -mixing is caused by the presence of Goldstone mode 
in the scalar leptoquark sector and this mixing is an intrinsic feature 
of the model. 
Our results (\ref{eq:slq})-(\ref{eq:ulq}) differ from those of Ref.\cite{LL} 
due to the account of the 
Goldstone mode and due to the more general form of the scalar leptoquark 
mixing.  

Secondly, in the case of the degenerated scalar leptoquark masses 
$ m_k=m_{-}=m_{+}\equiv m_S $ the expressions 
(\ref{eq:slq})-(\ref{eq:ulq}) give 
\begin{equation}
S^{(LQ)}=\frac{n_c\xi^2}{12\pi}[-12m_V^2\frac{f_1(m_S,m_V)}{(m_S^2-m_V^2)^2}+ 
f_2(m_S,m_V)+\ln\frac{m_V^2}{m_S^2} ]   \label{eq:sslq}
\end{equation} 
and
$$
T^{(LQ)}=0,U^{(LQ)}=0,
$$
whereas the contributions of type (\ref{eq:sp})-(\ref{eq:up}) into $S, T, U$ 
from the ordinary 
scalar doublets in the case of all degenerated masses are equal to zero.  
It should be noted, however, that the contribution (\ref{eq:sslq}) is small 
because 
of the smallness of the parameter $\xi$.  In the particular cases of 
$m_S\gg m_V, m_S=m_V,$ and $m_S\ll m_V$ the contribution (\ref{eq:sslq}) takes 
the values $S^{(LQ)}=-5n_c\xi^2/36\pi, S^{(LQ)}=-n_c\xi^2/3\pi $ and 
$S^{(LQ)}=-(n_c\xi^2/12\pi)(41/3+2\ln(m_V^2/m_S^2))$ 
respectively. 

The splitting of the scalar leptoquark masses varies the situation 
essentiallly. Because the scalar leptoquark masses are defined by 
Yukava coupling constants and by the VEV's including the large 
VEV's $\eta_1$ and $\eta_4$ their splitting can be rather large 
giving the possibility for $S$ and $T$  to take the large 
values, positive or negative in dependence on the mass splitting 
and the mixing parameters.

We have carried out the numerical analysis of the contributions 
(\ref{eq:slq})-(\ref{eq:ulq}) and (\ref{eq:sp})-(\ref{eq:up}) using 
the responsible for a new physics experimental 
values of  $S_{new}^{exp},T_{new}^{exp},U_{new}^{exp}$  \cite{PDG96}
\begin{equation}
S_{new}^{exp}= -0.28\pm 0.19,
T_{new}^{exp}= -0.20\pm 0.26,
U_{new}^{exp}= -0.31\pm 0.54 
\label{eq:stue1}        
\end{equation}
by minimizing $\chi^2$ defined as 
$$
\chi^2=\frac{(S-S_{new}^{exp})^2}{(\Delta S)^2}+
\frac{(T-T_{new}^{exp})^2}{(\Delta T)^2}+
\frac{(U-U_{new}^{exp})^2}{(\Delta U)^2},
$$
where $S,T,U$ are the leptoquark contributions (\ref{eq:slq})-(\ref{eq:ulq}) 
or the sum of the contributions (\ref{eq:slq})-(\ref{eq:ulq}) 
and (\ref{eq:sp})-(\ref{eq:up}) and $\Delta S,\Delta T, \Delta U$ 
are  the experimental errows in (\ref{eq:stue1}).

In following, we assume for simplicity that the fields 
$S_2^{(+)}$, and $\starup{S_2^{(-)}}$ in (\ref{eq:sup}) contain 
(in addition to 
the Goldstone mode $S_0$) only two physical fields $S_1$ and $S_2$,  
that is,we assume that $\theta_{13}=\theta_{23}=0$. 
In this case the contributions (\ref{eq:slq})-(\ref{eq:ulq}) 
depend on the vector leptoquark 
mass $m_V$, parameter $\xi$ and on the four scalar leptoquark masses 
$m_{+}=m_{5/3},m_{-}=m_{1/3},m_{1}=m_{2/3},m_{2}=m_{2/3}'$ 
and one mixing angle $\theta_{12}$. Here indexes $5/3, 1/3, 2/3 $ 
of the mass denote the electric charges of the corresponding scalar 
leptoquarks. 

The dependence of $S^{(LQ)},T^{(LQ)},U^{(LQ)}$ on the large $m_V$ 
(of order of hundreds TeV) is slight because of the smallness of 
the parameter $\xi$. For definiteness we assume further that 
$m_V\sim 100\,\, TeV$ and $\xi\sim 10^{-4}$. The numerical analysis 
of the contributions (\ref{eq:slq})-(\ref{eq:ulq}) by minimizing 
$\chi^2$ at these values 
of $m_V$ and $\xi$ shows that, firstly, $S^{(LQ)}, T^{(LQ)}, U^{(LQ)}$ 
and $\chi^2$ depend on the ratios of scalar leptoquark masses 
and, secondly, the value of the mixing angle $\theta_{12}=0$ 
is favored. $\chi_{min}^2$ as a functions of each of the independent 
mass ratios $\mu_{2/3}=m_{2/3}/m_{5/3}$,  $\mu_{1/3}=m_{1/3}/m_{5/3}$ 
and $\mu_{2/3}'=m_{2/3}'/m_{5/3}$ 
in the case of degenerated doublets 
$\Phi'$ and $F_j$ are shown in Fig.2
at $\theta_{12}=0$. Here we denote by $\chi_{min}^2(\mu)$
the minimal value of $\chi^2$ obtained at fixed value $\mu$ of one 
of the mass ratios by minimizing $\chi^2$ over two other mass ratios. 

It is seen from Fig.2 that experimental values (\ref{eq:stue1}) favor 
one scalar 
leptoquark $S_1$ with electric charge 2/3 to be the lightest one 
the scalar leptoquark $S_1^{(-)}$ with electric charge 1/3 to be slightly 
heavyer whereas the other scalar leptoquarks $S_2$ and $S_1^{(+)}$ 
with electric charges 2/3 and 5/3 respectively to be the 
heavyest ones.

The Fig.3 shows the regions in the plane  of the lightest masses 
$\mu_{2/3}$ and $\mu_{1/3}$ compatible with the data (\ref{eq:stue1}) 
at 95\% CL and 90\% CL ( the inner regions bounded by 
the solid and bold lines respectively ).

It is interesting to note that the scalar leptoquarks improve 
the agreement of the experimental values (\ref{eq:stue1}) with the theory 
so that $\chi^2$ can be reduced from the value $\chi^2=3.1$ 
(i.e. 38\% CL) of the SM to the value $\chi^2\approx 0.22$ 
(i.e. 97\% CL) in the model under consideration. For example, 
at the mass ratios near the values 
$\mu_{2/3}=0.01$,  $\mu_{1/3}=0.10$,  $\mu_{2/3}'=0.96 $
the formulas (\ref{eq:slq})-(\ref{eq:ulq}) give the values 
\begin{equation}
 S^{(LQ)}=-0.27, T^{(LQ)}=-0.20,  U^{(LQ)}=-0.06  
\label{eq:stum}
\end{equation} 
which agree with (\ref{eq:stue1}) with $\chi^2=0.22 $ (i.e. at 97\% CL). 
It should be noted, however, that attainment of the values (\ref{eq:stum}) 
demands a fine mutual fitting of the mass ratios near 
the values mentioned above as if there were some relation 
between the masses of the scalar leptoquarks. Possibly, 
such mass relation can arise as a result of the broken 
four-color symmetry implied here.

The account of the mass splitting of the doublets 
$\Phi'$ and $F_j$ modifies the result not essentially 
prefering slightly the down neutral components  of these 
doublets to be lighter than the up charged ones.
For example, at $m_{\Phi_2'}/m_{\Phi_1'}=0.5 $ the deviations 
of $\chi_{min}^2$ and of the boundaries of regions from those 
shown in Figs.2,3 do not exceed 10\%. It should be noted 
that with the existence of the salar leptoquark doublets 
the mass splittings of other doublets are allowed 
to be rather large. The arising in this case 
large positive contribution into $T$ from these doublets 
can be absorbed by the large negative contribution from 
the scalar leptoquark, which results in the values compatible 
with the experimental ones.

Recently a new analysis of the current electroweak data 
has been carried out in Ref.\cite{HHM}. Using the results of Ref.\cite{HHM} 
one can obtain 
\begin{equation}
S_{new}^{exp}= -0.11\pm 0.13,
T_{new}^{exp}= -0.03\pm 0.14,
U_{new}^{exp}=  0.03\pm 0.38  
\label{eq:stue2}     
\end{equation} 
at $m_t=175 \, GeV$, $m_{H^0}=300 \, GeV$. 

The contributions (\ref{eq:slq})-(\ref{eq:ulq}) have  
a large mass region compatible with 
these data at high CL. As an example the dotted curve in Fig.3 
shows the region in plane of $\mu_{2/3}$, $ \mu_{1/3}$ 
compatible with the data (\ref{eq:stue2}) at 99\% CL. 

It should be pointed out that both data (\ref{eq:stue1}) and 
(\ref{eq:stue2}) do not 
impose any lower limit on the mass $m_{2/3}$ of the 
scalar leptoquark $S_1$ with electric charge 2/3 
(see Figs.2,3), which gives the possibility for this 
mass (in contrast, for example, to $m_{5/3}$) to be light, 
perhaps to lie even not far from the SM mass scale. 
In this case the scalar leptoquark $S_1$ could manifest 
itself in experiments at a moderate energies, 
in particular, through the $e^+d\,\,\starup{S_1}$ - coupling 
in $e^+p$ - collisions.

In conclusion we resume the results of the work. 
The contributions into radiative correction parameters 
$S, T, U $ from the scalar leptoquarks are calculated 
in the minimal gauge model with four-color 
quark-lepton symmetry.
 
It is shown that the contributions 
into $T$ and $U$ from the scalar leptoquark doublets are 
not positive definite due to the mixing between the 
scalar leptoquark fields with electric charge 2/3, 
which is caused by the existence of the Goldstone mode 
among these fields.
 
Using the current experimental limits on $S, T, U $ 
the numerical analysis of these contributions is 
carried out and the bounds on the scalar leptoquark 
masses are obtained.In particular,it is shown that 
the current experimental data on $S, T, U $ allow 
the existence of the relatively light scalar leptoquark 
with electric charge 2/3 .

\bigskip

{\bf Acknowledgment}

\bigskip

The author is grateful to A.V.~Povarov for the help in the work.

\newpage

{\Large\bf Figure captions}

\bigskip

\begin{quotation}

\noindent 
Fig. 1. The self energy diagramms contributing into $S$, $T$, $U$ 
        from the scalar leptoquarks $ S_1^{(\pm)}$, $S_k$, $k= 1, 2, 3 $ 
        ($V$ is the vector leptoquark, $X$, $Y$ are 
        the photon and $Z$-boson). \\

\noindent 
Fig. 2. $\chi_{min}^2$ as a function of the scalar leptoquark 
        mass ratios $\mu_{2/3}=m_{2/3}/m_{5/3}$  (the solid line), 
        $\mu_{1/3}=m_{1/3}/m_{5/3}$  (the bold line),    
        and $\mu_{2/3}'=m_{2/3}'/m_{5/3}$  (the dotted line). \\
 
\noindent 
Fig. 3. The regions in plane of mass ratios 
        $\mu_{2/3}=m_{2/3}/m_{5/3}$, 
        $\mu_{1/3}=m_{1/3}/m_{5/3} $ compatible with the data~(\ref{eq:stue1}) 
        at 95\% CL (the solid line), at 90\% CL (the bold line) and with 
        the data~(\ref{eq:stue2}) at 99\% CL (the dotted line).

\end{quotation}

\newpage

\begin{figure}[htb] 

\epsffile[110 450 0 700]{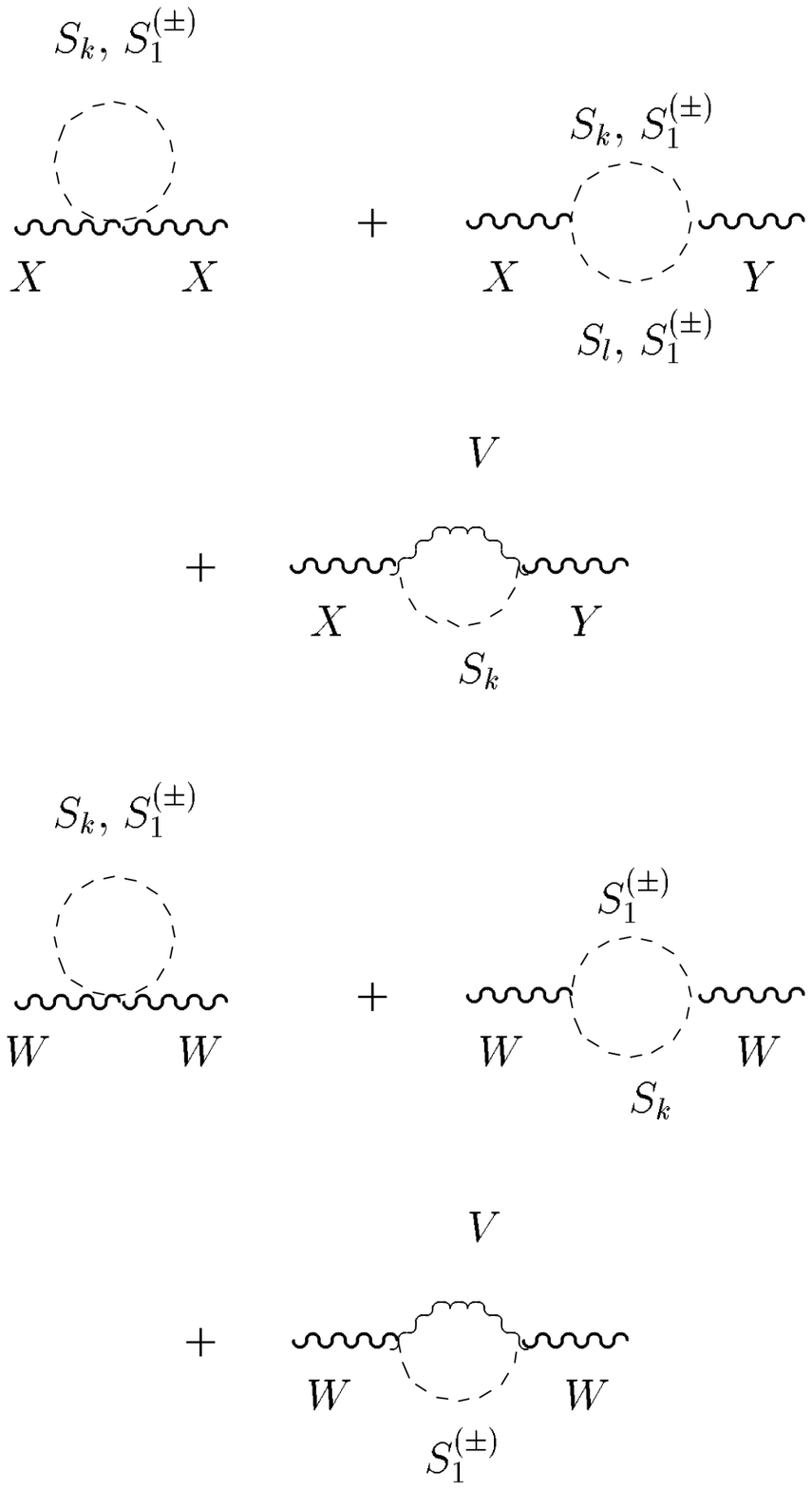}

\end{figure}

\vfill
\centerline{A.D.~Smirnov, Physics Letters B}

\centerline{Fig. 1}

\newpage

\begin{figure}[htb] 

\epsffile[110 450 0 700]{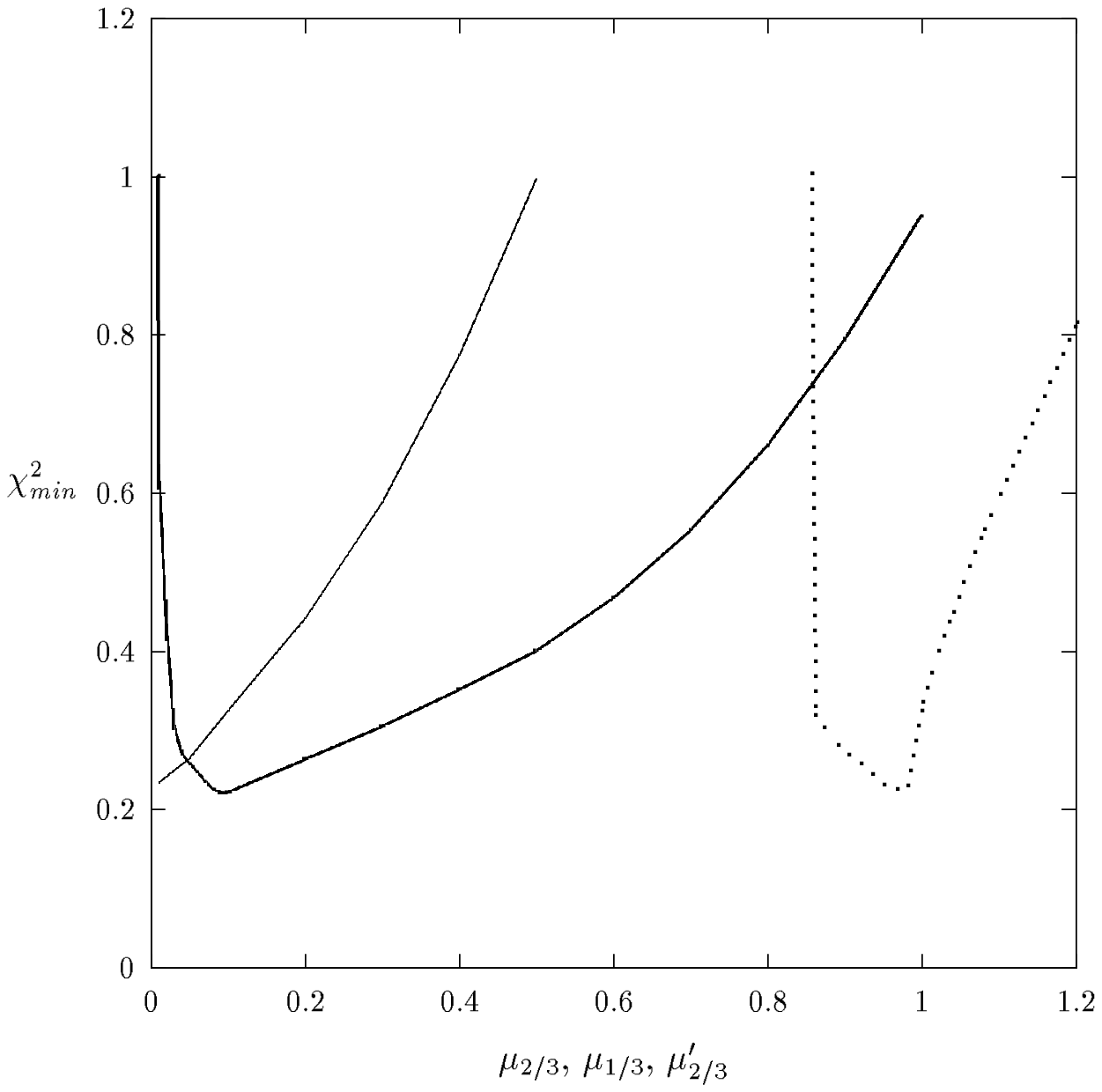}

\end{figure}

\vfill
\centerline{A.D.~Smirnov, Physics Letters B}

\centerline{Fig. 2}

\newpage

\begin{figure}[htb] 

\epsffile[110 450 0 700]{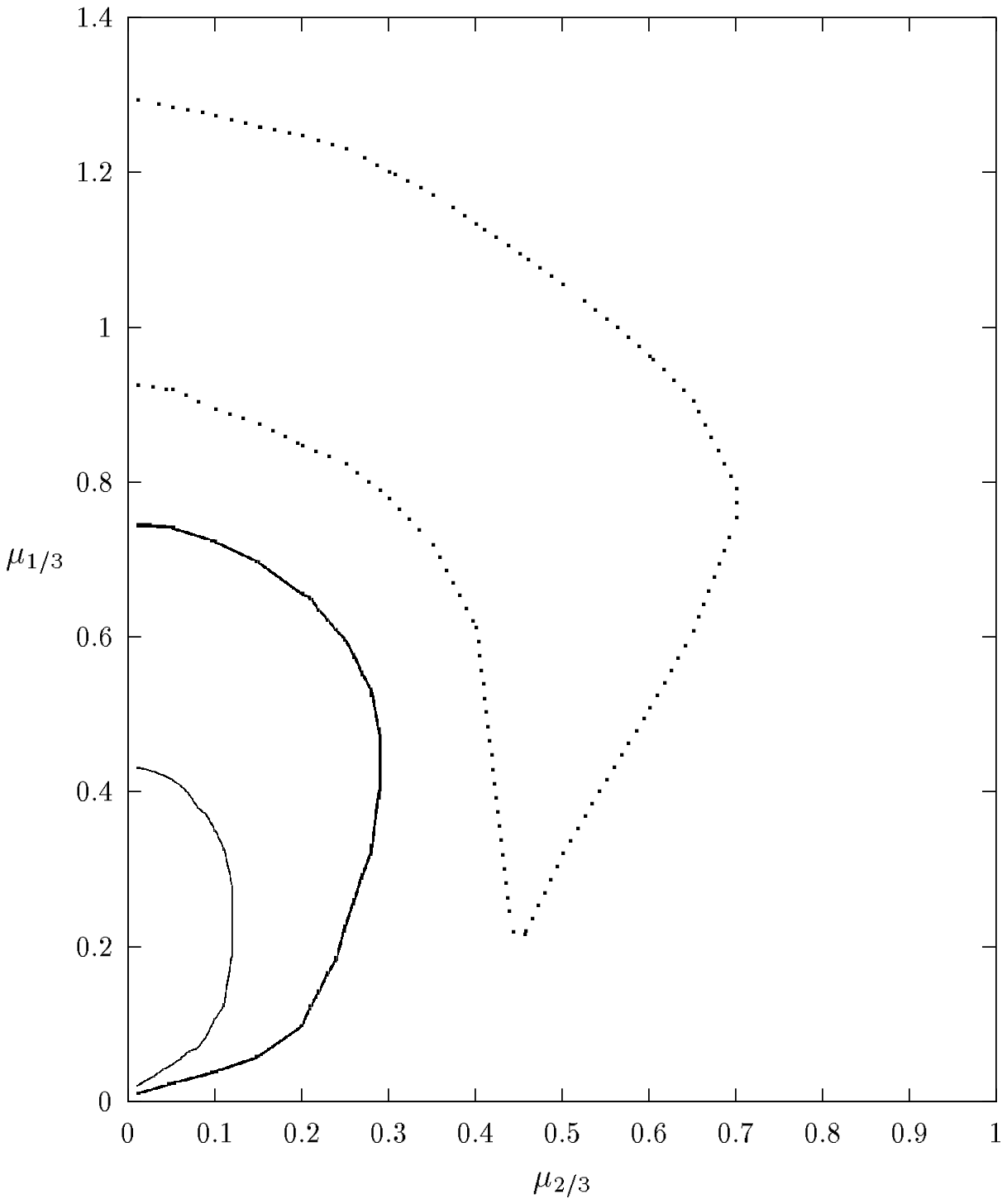}

\end{figure}

\vfill
\centerline{A.D.~Smirnov, Physics Letters B}

\centerline{Fig. 3}


\newpage

\begin{thebibliography}{7}

\bibitem{PS}
   J.C.~Pati and A.~Salam, Phys.~Rev. D10~(1974)~275.
\bibitem{EF}
   E.M.~Freire, Phys.~Rev. D43~(1991)~209.
\bibitem{SS}
   G.~Senjanovic, A.~Sokorac, Z.Phys. C20~(1983)~255.
\bibitem{AD1}
   A.D.~Smirnov, Phys.~Lett. B346~(1995)~297.
\bibitem{AD2}
   A.D.~Smirnov, Yad.~Fiz. 58~(1995)~2252,
                 Phys. of At. Nucl. 58~(1995)~2137.
\bibitem{V}
   R.R.~Volkas, Phys.~Rev. D53~(1996)~2681.
\bibitem{RF}
   R.~Foot, UM-P-97/48, RCHEP-97/08, hep-ph/9708205. 
\bibitem{BL}
   A.~Blumhofer, B.~Lampe, MPi-PhT/97-37, LMU-07/97, hep-ph/9706454. 
\bibitem{PT}
   M.E.~Peskin and T.~Takeuchi, Phys.~Rev. D46~(1992)~381.
\bibitem{LL}
   L.~Lavoura and L.F.~Li, Phys.~Rev. D48~(1993)~234. 
\bibitem{PDG96}
   Particle Data Group, R.M.~Barnet et~al., Phys.~Rev. D54~(1996) 
   part~1,~7.
\bibitem{HHM}
   K.~Hagivara, D.~Haidt, S.~Matsumoto, DESY preprint DESY 96-192 (1997),
   KEK preprint KEK-TH-512(1997), hep-ph/9706331.



\end{thebibliography}
\end{document}